\documentclass[sigconf]{acmart}
\usepackage{algorithm}
\usepackage{algpseudocode}
\usepackage{subfigure}
\usepackage{svg}
\usepackage{multirow}
\usepackage{booktabs}
\usepackage{siunitx}

\AtBeginDocument{%
  \providecommand\BibTeX{{%
    \normalfont B\kern-0.5em{\scshape i\kern-0.25em b}\kern-0.8em\TeX}}}

\copyrightyear{2022}
\acmYear{2022}
\setcopyright{acmcopyright}\acmConference[MMAsia '22]{ACM Multimedia Asia}{December 13--16, 2022}{Tokyo, Japan}
\acmBooktitle{ACM Multimedia Asia (MMAsia '22), December 13--16, 2022, Tokyo, Japan}
\acmPrice{15.00}
\acmDOI{10.1145/3551626.3564965}
\acmISBN{978-1-4503-9478-9/22/12}

\begin{document}

\title{A Multimodal Sensor Fusion Framework Robust to Missing Modalities for Person Recognition}

\author{Vijay John}
\email{vijay.john@riken.jp}
\affiliation{%
  \institution{Guardian Robot Project, RIKEN}
  \country{Japan}
}

\author{Yasutomo Kawanishi}
\email{yasutomo.kawanishi@riken.jp}
\affiliation{%
  \institution{Guardian Robot Project, RIKEN}
  \country{Japan}
}


\begin{abstract} 
Utilizing the sensor characteristics of the audio, visible camera, and thermal camera, the robustness of person recognition can be enhanced.
Existing multimodal person recognition frameworks are primarily formulated assuming that multimodal data is always available.
In this paper, we propose a novel trimodal sensor fusion framework using the audio, visible, and thermal camera, which addresses the missing modality problem.
In the framework, a novel deep latent embedding framework, termed the AVTNet, is proposed to learn multiple latent embeddings.
Also, a novel loss function, termed missing modality loss, accounts for possible missing modalities based on the triplet loss calculation while learning the individual latent embeddings.
Additionally, a joint latent embedding utilizing the trimodal data is learnt using the multi-head attention transformer, which assigns attention weights to the different modalities.
The different latent embeddings are subsequently used to train a deep neural network.
The proposed framework is validated on the \textit{Speaking Faces} dataset.
A comparative analysis with baseline algorithms shows that the proposed framework significantly increases the person recognition accuracy while accounting for missing modalities.
\end{abstract}


\maketitle

\section{Introduction}

\begin{figure*}[!ht]
 \begin{center}
\subfigure[]{\includegraphics[width=45mm]{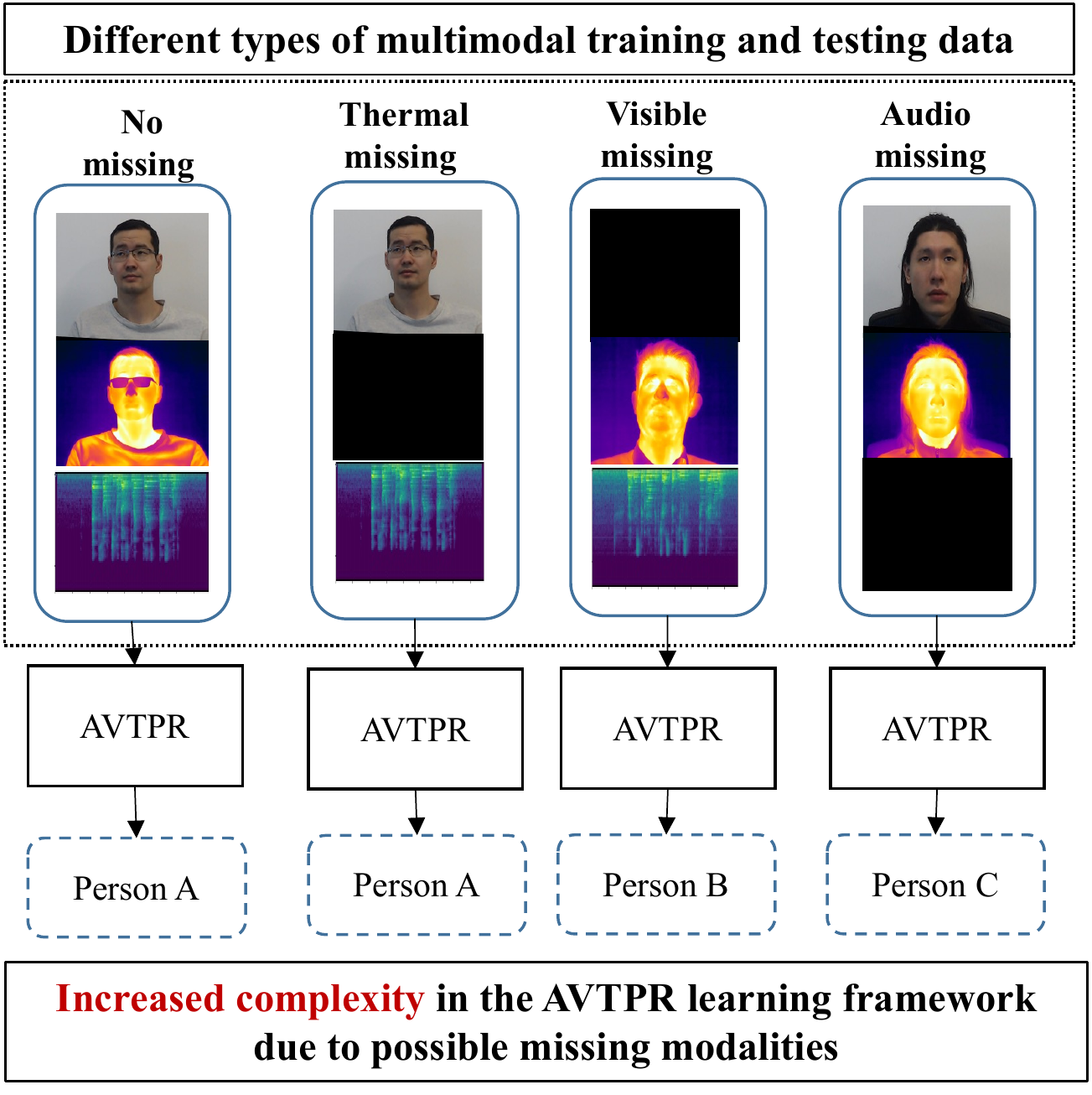}} \hspace{0.5cm}
\subfigure[]{\includegraphics[width = 118mm]{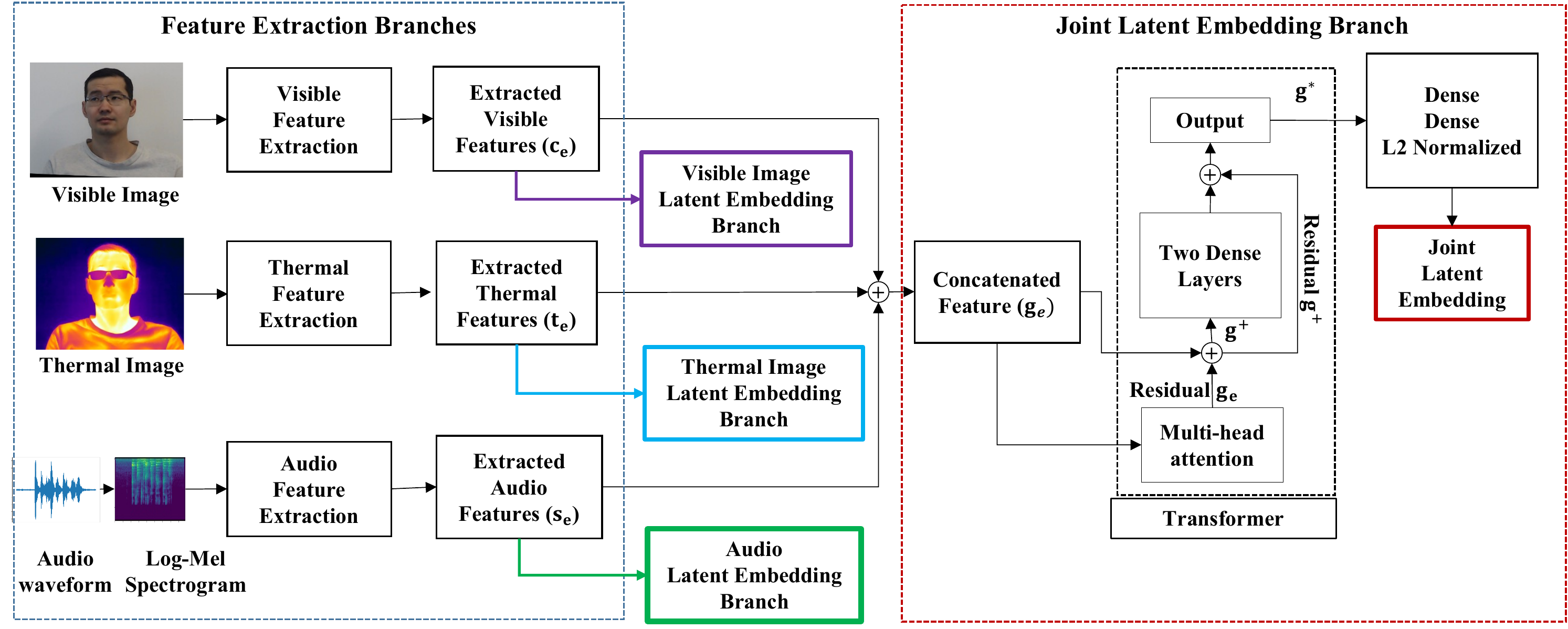}}
 \end{center}
  \caption{(a) An illustration of the missing modality problem in multimodal person recognition. (b) An overview of the AVTNet}
  \label{fig:multimodal}
\end{figure*}

Audio-visible person recognition (AVPR)~\cite{Seyed, Das, Tao, Nagrani} and thermal-visible person recognition (TVPR)~\cite{Kanmani, Seal, Kong,  Bebis, Saurabh} report enhanced recognition accuracy owing to the complementary characteristics of the different sensors. The performance of the AVPR and TVPR can be further enhanced using an audio-visible-thermal camera person recognition (AVTPR) framework by utilizing the sensor characteristics of all three sensors.

Existing works in audio-video and thermal-video fusion are formulated with the assumption that all the input modalities are present during training and inference. However, in real-world applications, there is the possibility of missing modalities due to conditions such as sensor malfunction or failure, resulting in missing data (Fig~\ref{fig:multimodal}~(a)). Under these circumstances, the performance of existing deep fusion frameworks is affected.

In this paper, we propose an AVTPR framework that addresses the missing modality problem. The proposed AVTPR framework consists of a deep latent embedding learning framework, termed AVTNet, which simultaneously performs multimodal sensor fusion while learning multiple latent embeddings.
To account for the missing modalities, the AVTNet utilizes a loss function strategy, termed missing modality loss, and the transformers~\cite{Vaswani} to learn multiple latent embeddings. 

The AVTNet learns four embeddings from the multimodal data, represented by three \textit{modal-specific} embeddings and one joint multimodal embedding. The missing modality loss is used to learn the three \textit{modal-specific} embeddings. The joint multimodal embedding learns the joint latent representation of the visible, thermal, and audio features using the multi-head attention-based transformer~\cite{Vaswani}. By utilizing the attention mechanism, the attention weights account for any missing modality while learning the joint embedding. The four learnt embeddings are used to train a deep learning-based person recognition model.

The proposed framework is validated on the Speaking Faces public dataset~\cite{SpeakingFaces}. A comparative analysis is performed with baseline algorithms, and a detailed ablation study is performed. The results show that the proposed framework addresses the missing modality problem.

The main contributions to literature are as follows:
\begin{itemize}
    \item A novel computer vision application framework, termed the AVTNet, for the visible camera, thermal camera, and audio-based person recognition. The AVTNet, addresses the missing modality problem.
    \item We introduce a tailored loss function strategy termed the missing modality loss, which learns the individual modal-specific embeddings while accounting for missing modalities. 
\end{itemize}

The remainder of the paper is structured as follows. The literature is reviewed in Section~\ref{sec:review}. The proposed framework is presented in Section~\ref{sec:algo}, and the experimental results are presented in Section~\ref{sec:experiments}. Finally, we summarize the research in Section~\ref{sec:conclusion}.

\section{Literature Review}
\label{sec:review}


In literature, the thermal-visible~\cite{Kanmani, Seal, Kong,  Bebis, Saurabh} (TVPR) and audio-visible~\cite{Wen, Nawaz, Li, Sell, Choudhury, Chetty, Tao} (AVPR) person recognition approaches report enhanced recognition accuracy owing to the complementary sensor characteristics of the different sensors. It can be observed that the complementary characteristics of the three sensors can be fused for further enhancement of the person recognition accuracy using a trimodal person recognition framework. Currently, to the best of our knowledge, there is no previous work in the person recognition literature utilizing the three sensors: the visible camera, the thermal camera, and the audio (AVTPR).

In recent years, several researchers have sought to address the missing modality problem~\cite{Ma,zhao} using two methods for various perception tasks. In the first method, a joint multimodal representation is learnt from the different modalities to overcome the possibility of missing modalities~\cite{Pham,Han, Zilong}. Han et al.~\cite{Han} propose an audio-visual emotion recognition framework that learns the joint audio and video representation. The learnt joint representation is used with a shared classifier addressing the missing modality problem. In the second method, a data augmentation strategy is adopted, where subsets of input modalities are randomly ablated during training to resemble real-world missing modality conditions. Training with randomly ablated data is shown to improve the recognition accuracy for the missing modality condition~\cite{Parthasarathy}. Parthasarathy et al.~\cite{Parthasarathy} address the missing modality problem for emotion recognition by randomly ablating the video or audio data during training. 

Compared to the literature, our proposed framework is the first AVTPR framework to address the missing modality problem. Our proposed approach to address the missing modality problem involves ablating the training data as well as learning multiple latent embeddings from the multimodal data.

\section{Algorithm}
\label{sec:algo}



The proposed framework robustly performs person recognition by learning multiple latent embeddings from the missing modality multimodal data using the AVTNet. The learnt embeddings are used to train a deep learning-based person recognition model. The latent embeddings are learnt from ablated training data, which reflect the conditions of missing modality.

\subsection{AVTNet}

The AVTNet is formulated as an End-to-End (E2E) deep learning framework which simultaneously learns the individual and joint deep latent spaces from the ablated audio, visible camera, and thermal camera data. The AVTNet contains three independent feature extraction branches, three independent latent embedding branches, and one joint latent embedding branch. An overview of the architecture is presented in Fig~\ref{fig:multimodal}~(b).

\subsubsection{Feature Extraction Branches}

\paragraph{Audio Branch:}  

Log-Mel-spectrogram ($\mathbf{s}$) is computed from the audio input. The audio feature extraction can be represented by the function $\mathbf{s_e}$=$f_s(\mathbf{s})$. The function $f_s$ is represented using multiple deep learning Conv1D layers. The details of the deep learning layers are presented in Section~\ref{ssec:algoparams}.

\paragraph{Visible and Thermal Camera Branches:} 

 Apart from audio, the visible colour camera image, $\mathbf{c}\in\mathbb{R}^{(224\times224\times3)}$, and thermal camera image, $\mathbf{t}\in\mathbb{R}^{(224\times224)}$, are also given as input to the AVTNet. The visible camera and thermal feature extractions are represented by the functions $\mathbf{c_e}$=$f_c(\mathbf{c})$ and  $\mathbf{t_e}$=$f_t(\mathbf{t})$. The functions $f_c$ and $f_t$ are represented using multiple deep-learning Conv2D and max pooling layers presented in Section~\ref{ssec:algoparams}.

\subsubsection{Individual Latent Embedding Branches:} Deep latent embeddings of the trimodal features are learnt individually using a tailored loss function, the missing modality loss. The missing modality loss function, based on the triplet hard loss, explicitly accounts for the missing modalities.

Here, the extracted features, $\mathbf{s_e}$, $\mathbf{c_e}$, and $\mathbf{t_e}$ are projected using two dense layers with $256$ neurons with ReLU and linear activation functions. The layer output is L2 normalized and given as input to the missing modality loss function. Details of the missing modality loss are presented in Section~\ref{sssec:missingmodalloss}.

\subsubsection{Joint Latent Embedding Branch:} The joint latent embedding of the multimodal data is learnt from the concatenation of the three extracted features, $\mathbf{g_e}$ = [$\mathbf{s_e}$, $\mathbf{c_e}$, $\mathbf{t_e}$], using the transformer model and the triplet hard loss. The transformer utilizes multi-head attention to compute the self-attention for the concatenated feature $\mathbf{g_e}$. The self-attention inherently accounts for any missing modality using the attention weights. The multi-head attention computes the attention using four heads. The architecture of the transformer is shown in Fig~\ref{fig:multimodal}~(b). The two dense layers in the transformer branch have $400$-$64$ neurons and ReLU-linear activations. The transformer output, $\mathbf{g}^*$, is given as an input to two dense layers with $256$ neurons with ReLU and linear activation functions. The dense layer output is L2 normalized and given as input to the triplet hard loss function. 

\subsection{Loss Functions for Latent Embeddings}
The overall loss function for the AVTNet framework is represented as,

\begin{equation}
\label{eqn:losses}
    \mathcal{L} = \mathcal{L}_c + \mathcal{L}_t + \mathcal{L}_s + \mathcal{L}_j
\end{equation}

where $\mathcal{L}_c$, $\mathcal{L}_t$, and $\mathcal{L}_s$ correspond to the missing modality losses for the modal-specific latent embedding. $\mathcal{L}_j$ corresponds to the triplet hard loss for the joint latent embedding. 

\begin{figure}[!t]
 \begin{center}
\includegraphics[width=85mm]{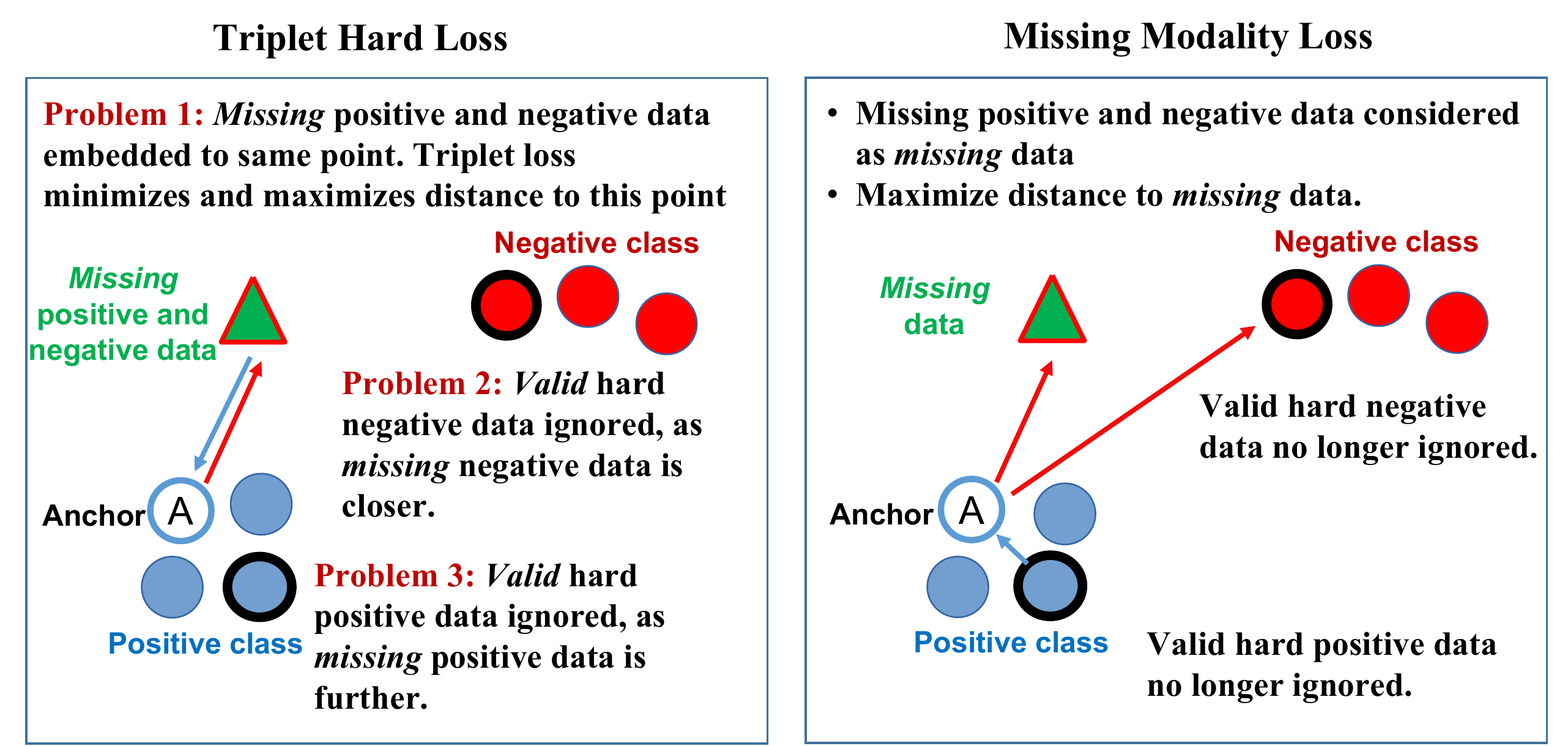}
 \end{center}
  \caption{An illustration of the triplet hard loss and the missing modality loss}
  \label{fig:loss}
\end{figure}





The triplet hard loss is optimized with the assumption that the sensor data is always present without accounting for missing data. If the missing data is naively represented using pre-defined fixed data, there is a possibility that the fixed data is considered as either an anchor, positive or negative data within the identified triplets. 

Additionally, note that the pre-defined fixed data corresponding to the missing modality are embedded to the same latent point, $e^m$. Thus, for a triplet with valid anchor, missing positive, and missing negative latent points, the triplet loss tries to simultaneously minimize and maximize the distance between the valid anchor point and the missing positive and negative points. Since the missing points are embedded to the same latent point ($e^m$), the triplet loss learns a sub-optimal latent space in the presence of missing modality data. The missing modality loss is proposed to address this limitation. An illustration is shown in Fig~\ref{fig:loss}.


\subsubsection{Individual Embedding: Missing Modality Loss}
\label{sssec:missingmodalloss}

Each missing modality loss in Equation~(\ref{eqn:losses}), $\mathcal{L}_z, z \in \{s,c,t\}$, for the audio, visible camera, and thermal camera is given as
\begin{align} 
\mathcal{L}_z & = log_e(1 + e^\alpha), \\ 
\alpha  & = d(a^v,p^v) - d(a^v,n^v) -  d(a^v,e^m).
\label{eqn:components}
\end{align}

$d(a^v,p^v)$ represents the Euclidean distance between \textit{valid} anchor points and the hard \textit{valid} positive points. $d(a^v,n^v)$, represents the Euclidean distance between \textit{valid} anchor points and the hard \textit{valid} negative points. $d(a^v,e^m)$ represents the Euclidean distance between \textit{valid} anchor points and the \textit{missing} latent points.  

Here \textit{valid} points correspond to data obtained from functional sensors. While \textit{missing} corresponds to the missing modality pre-defined fixed data. The hard positive corresponds to the similar class data farthest from the anchor. The hard negative corresponds to the dissimilar class data, which is nearest to the anchor. 

To compute the loss in $\mathcal{L}_z$, the training dataset is partitioned into mini-batches with $K$ samples each. Each mini-batch contains input features $\mathbf{X}=\{\mathbf{x}_i\}_{i=1}^K$, class labels, $\mathbf{Y}=\{y_i\}_{i=1}^K$, and binary validity label $\mathbf{B}=\{b_i\}_{i=1}^K$. Here, $b_i$=$1$ corresponds to \textit{valid} features, and $b_i$=$0$ corresponds to the \textit{missing} features. 

To compute the loss, firstly, the pairwise distance matrix $P$ is computed over the features $\mathbf{X}$. Each row in $P$ represents the distance vector between an anchor point and all other data points in the mini-batch. 

Next, multiple 2D mask matrices, the positive matrix $A_{p}$, the \textit{valid} matrix $A_{v}$, and the missing matrix $A_{m}$ , are computed using the class labels, $\mathbf{Y}$, and binary validity labels, $\mathbf{B}$. 

\begin{align}
A_{p}(i,j) &= 
    \begin{cases}
     1, & \text{if }  y_i = y_j \\
     0, & \text{otherwise}
    \end{cases}\\
A_{v}(i,j) &= 
    \begin{cases}
     1, & \text{if }  b_i = 1 \land b_j = 1 \\
     0, & \text{otherwise}
    \end{cases}\\
A_{m} &= \lnot A_{v}
\end{align}

Next, the \textit{valid} positive matrix, $A_{p}^v$, the negative matrix $A_{n}$, and the \textit{valid} negative matrix $A_{n}^v$ are computed. The diagonal elements of $A_{p}^v$ are set to zero.

\begin{align}
 A_{p}^v & = A_{p} \land A_{v} \\
 A_{n}    & = \lnot A_{p}  \\
 A_{n}^v  & = A_{n} \land A_{v} 
\end{align}

\noindent \textit{Loss Components:} In Equation~(\ref{eqn:components}), the $d(a^v,p^v)$ component is computed by the following steps. Firstly, the Hadamard product between $P$ and the mask matrix $A_{p}^v$ is performed to obtain $P_{p}^v$. In $P_{p}^v$, a row-wise maximum and an element-wise multiplication with $\mathbf{B}$ are performed to generate the first loss component.

Similarly, $d(a^v,n^v)$ and $d(a^v,e^m)$, are computed by the Hadamard product between $P$ and $A_{n}^v$ and $A_{m}$ to obtain $P_{n}^v$ and $P_{m}$, respectively, In these matrices, a row-wise minimum and an element-wise multiplication with $\mathbf{B}$ are performed to generate the two loss components. 

\subsubsection{Joint Embedding: Triplet Hard Loss}
The joint latent embedding is learnt from the concatenated feature; in the current work, there are no conditions where all the three modalities are missing. Thus, we use the triplet hard loss $\mathcal{L}_j$ to learn the joint latent embedding.

\subsection{Person Recognition Model}

The four learnt latent embeddings are concatenated and given as input to the deep learning-based person recognizer. The person recognizer consists of two dense layers with $512-256$ units, batch normalization, and ReLU activation function. The layer output is given to the output layer with $75$ neurons and softmax activation function. The deep learning model is trained with sparse categorical cross-entropy.

\section{Experiments} 
\label{sec:experiments}

We validate the proposed framework using the public Speaking Faces dataset~\cite{SpeakingFaces}. Comparative analysis with baseline algorithms and ablation study is performed. 

\subsection{Dataset and Algorithm Parameters}
\label{ssec:algoparams}

We selected $3,473$ synchronized audio-visible-thermal video samples from $75$ people in the speaking dataset. For each selected synchronized sample, we generated three ablation samples with missing modalities to obtain a dataset of $13,893$ samples. In the first ablation sample, the audio sequence's log-Mel spectrogram image of size $128 \times 589$ is filled with zero. Similarly, in the second and third ablation samples, the visible and thermal images are represented by zero images of size $224 \times 224 \times 3$ and $224 \times 224 \times 1$, respectively. The evaluation dataset with ablation samples is randomly partitioned into a training dataset with $11,164$ samples and a test dataset with $2,729$ unknown samples. 


The audio feature extraction ($f_s$) is performed using three layers of Conv-1D with $64$ filters of size $11$ with stride $1$ and ReLU activation. The visible and thermal feature extractions ($f_c$, $f_t$) are performed using four layers of Conv-2D filters with stride $1\times1$ and ReLU activation. Each of the three Conv-2D layers contains $128$ filters with filter size $3\times3$ followed by a max-pooling 2D layer with pooling size $2\times2$. The final Conv-2D layer contains $64$ filters with filter size $1\times1$.

From each video sample, the audio spectrogram was calculated from the audio with a sampling rate of \SI{44,000}{\hertz}. The AVTNet was trained for $50$ epochs with a batch size of $32$ from which the missing modality and triplet online mining were performed. The trained AVTNet generates the training and testing embedded data to train the person recognition model. The person recognition model is trained for $25$ epochs with batch size $32$. Both of the phases are trained with the ADAM optimizer with a learning rate of $0.001$, $\beta_1$=$0.5$, and $\beta_2$=$0.99$. The algorithm was implemented with Tensorflow 2 using NVIDIA $3090$ GPUs on an Ubuntu 20.04 desktop.  

\subsection{Baseline Algorithms and Ablation Study}



        

        
        


\begin{table}[!t]
\scriptsize
\centering
\caption{Comparative analysis of the recognition accuracy.}
\begin{tabular}{llllll}
        \toprule

\textbf{Algorithm}  &  {No-Missing} &  {Miss. Visible} & {Miss. Thermal} & {Miss. Audio} & {Avg.} \\
\midrule
\textbf{(Prop)} & \textbf{100} & 99.71 & \textbf{97.94} & \textbf{99.86} & \textbf{99.39}\\

\textbf{(JER-1)} & \textbf{100}  & 98.16 & 95.15 & 99.85 & 98.31 \\

 \textbf{(JER-2)} & 99.85  & 94.92 & 91.33 & 98.31 & 96.13 \\

 \textbf{(E2E)} & 99.70 & 90.70 & 89.5 & 99.80 & 94.90 \\

  \textbf{Dense-Triplet} & 99.85 & 99.71 & 87.37 & 97.33 & 96.13 \\
  \textbf{(Prop-I)} & \textbf{100}  & 99.57 & 95.88 & 98.45 & 98.49 \\
 
 \textbf{(Prop-II)} & \textbf{100}  & \textbf{100} & 91.04 & 93.68 & 96.20 \\
 \textbf{(Prop-III)} & \textbf{100}  & 93.93 & 92.65 & 99.43 & 96.52 \\
\bottomrule
\end{tabular}
\label{table:comparative}
\end{table}

End-to-End (E2E): This model contains three feature extraction branches followed by the recognition branch. The three feature extraction branches are similar to the AVTNet's feature extraction branches. The concatenated feature maps are given as input to the recognition branch, which contains three dense layers with $512-256-75$ neurons with ReLU, ReLU, and softmax activations, respectively. The first two layers have batch normalization. 

Joint embedding recognizer (JER-1 and JER-2): A single joint embedding is learnt using the three dense layers using the triplet loss (JER-1) and the triplet prototypical loss (JER-2). The concatenated feature maps are given as input to three dense layers with $256$ neurons with ReLU, ReLU, and linear activation functions. The learnt joint embedding is used to train the deep learning-based person recognition model. 

Dense-Triplet: Here, the three latent embeddings in the AVTNet are learnt with the triplet hard loss. Additionally, the joint embedding is learnt using three dense layers with $256$ neurons with ReLU, ReLU, and linear activation functions, respectively, instead of the transformer.

Prop-I, Prop-II, and Prop-III are the three variants of the proposed framework. In Prop-I, the concatenated feature maps are given as input to three dense layers with $256$ neurons with ReLU, ReLU, and linear activation functions, respectively. In Prop-II, all the latent embeddings are learnt using the triplet hard loss. In Prop-III, only the joint embedding is used.

We also perform a comparison with three bimodal variants. The architectures for the bimodal variants are similar to the proposed framework except for the absence of the third modality's feature extraction and latent embedding branches.

\begin{table}[!t]
\scriptsize
\centering
\caption{Accuracy of varying sensor fusion frameworks.}
\begin{tabular}{llll}
        \toprule
Proposed (Trimodal)  &  Audio-Visible &  Audio-Thermal  & Visible-Thermal  \\
\midrule
\textbf{99.39} & 95.62 & 92.53 & 98.52 \\
 \bottomrule
\end{tabular}
\label{table:sensorfusion}
\end{table}


 

\subsection{Discussion}

The results show that the proposed framework reports the best results (Table~\ref{table:comparative}). Comparing the proposed framework with \textit{Prop-I} and \textit{Prop-II} clearly show the advantages of utilizing the transformers and the novel loss function. The comparison with \textit{Prop-III} shows the advantages of learning multiple embeddings instead of the joint embedding alone. Results in Table~\ref{table:sensorfusion} demonstrate the advantage of trimodal sensor fusion.


\section{Conclusion} 
\label{sec:conclusion} 
In this paper, we propose an audio-visible-thermal person recognition framework (AVTPR) to address the missing modality problem. The AVTPR framework contains the AVTNet, which learns multiple embeddings using the missing modality loss and the multi-head attention-based transformer. The learnt embeddings are used to train a deep neural network-based recognizer. The proposed framework is validated on the Speaking Faces dataset, and the results show that the proposed framework addresses the missing modality problem. In the future, we will evaluate the proposed framework under conditions of multiple missing modalities.
\balance
\bibliography{z2022acmmm}
\bibliographystyle{ACM-Reference-Format}
\end{document}